\documentclass[5p,times,twocolumn,final]{elsarticle}

\usepackage{amssymb}
\usepackage{latexsym}

\usepackage{url}
\usepackage[table]{xcolor}
\definecolor{newcolor}{rgb}{.8,.349,.1}

\usepackage[pdftex,colorlinks]{hyperref}

\journal{Computers \& Graphics}

\usepackage[utf8]{inputenc}
\usepackage[T1]{fontenc}
\usepackage{algorithmic}
\usepackage[section]{algorithm}
\usepackage{amsmath,amssymb,amsfonts,latexsym,cancel}
\usepackage{amsthm}
\usepackage{tikz}
\usepackage{subcaption}
\usepackage{xstring}
\usepackage{thmtools,thm-restate}
\usepackage{xspace}
\usepackage{comment}

\makeatletter
\AtBeginDocument{\def\@citecolor{cyan!64!black}}
\AtBeginDocument{\def\@linkcolor{cyan!64!black}}
\AtBeginDocument{\def\@urlcolor{cyan!64!black}}
\makeatother

\def\appname{Parallel Toplesets Propagation\xspace}
\def\appabrv{PTP\xspace}
\def\topleset{\textit{topleset}\xspace}
\def\toplesets{\textit{toplesets}\xspace}

\def\image#1#2#3%
{%
	\begin{figure}[!htb]
		\centering
		\includegraphics[width=#2\linewidth]{#1}
		\caption{#3}\label{#1}
	\end{figure}
}

\def\simages#1#2#3#4
{
	\begin{figure}[!htb]
		\centering
		\foreach \imagepath/\imagename/\imagecaption in #2
		{
			\subcaptionbox{\label{\imagepath/\imagename} \em \imagecaption}
			{\includegraphics[width=#1\linewidth]{\imagepath/\imagename}}
		}
		\caption{#4}\label{#3}
	\end{figure}
}

\def\images#1#2#3#4
{
	\begin{figure*}[!htb]
		\centering
		\foreach \imagepath/\imagename/\imagecaption in #2
		{
			\subcaptionbox{\label{\imagepath/\imagename} \em \imagecaption}
			{\includegraphics[width=#1\linewidth]{\imagepath/\imagename}}
		}
		\caption{#4}\label{#3}
	\end{figure*}
}

\def\myimage#1#2%
{%
	\begin{figure}[!htb]
		\centering
		\input{#1}
		\caption{#2}\label{#1}
	\end{figure}
}

\def\mybimage#1#2%
{%
	\begin{figure*}[!htb]
		\centering
		\input{#1}
		\caption{#2}\label{#1}
	\end{figure*}
}

\def\myalgorithm#1#2
{
\begin{algorithm}[!htb]
	\caption{#2}\label{#1}
	\input{#1}
\end{algorithm}
}

\def\mytable#1#2
{
\begin{table}[!htb]
	\caption{#2}\label{#1}
	\centering\input{#1}
\end{table}
}

\def\mybtable#1#2
{
\begin{table*}[!htb]
	\caption{#2}\label{#1}
	\centering\input{#1}
\end{table*}
}

\newtheorem{theorem}{Theorem}
\newtheorem{lemma}[theorem]{Lemma}
\newtheorem{definition}{Definition}

\begin{document}


\begin{frontmatter}

\title{A minimalistic approach for fast computation of geodesic distances on triangular meshes}

\author[1,2,3]{Luciano A. {Romero Calla}}
\ead{romero@ifi.uzh.ch}
 
\author[1,2,3]{Lizeth J. {Fuentes Perez}}
\ead{fuentes@ifi.uzh.ch}

\author[2]{Anselmo A. {Montenegro}}
\ead{anselmo@ic.uff.br}

\address[1]{Visualization and MultiMedia Lab (VMML), Department of Informatics, University of Zurich, Switzerland}
\address[2]{Institute of Computing, Federal Fluminense University, Niteroi, Brazil}
\address[3]{IPRODAM3D research group, La Salle University, Arequipa, Peru}


\begin{abstract}

The computation of geodesic distances is an important research topic in Geometry Processing and 3D
Shape Analysis as it is a basic component of many methods used in these areas. In this work, we
present a minimalistic parallel algorithm based on front propagation to compute approximate geodesic
distances on meshes. Our method is practical and simple to implement, and does not require any heavy pre-processing.  The convergence of our algorithm depends on the number of discrete level sets around
the source points from which distance information propagates. To appropriately implement our method 
on GPUs taking into account memory coalescence problems, we take advantage of a graph representation
based on a breadth-first search traversal that works harmoniously with our parallel front propagation
approach. We report experiments that show how our method scales with the size of the problem. We
compare the mean error and processing time obtained by our method with such measures computed using
other methods. Our method produces results in competitive times with almost the same accuracy,
especially for large meshes. We also demonstrate its use for solving two classical geometry
processing problems: the regular sampling problem and the Voronoi tessellation on meshes.

\end{abstract}


\end{frontmatter}



\section{Introduction}\label{sec:introduction}

Computing geodesic distances on meshes is important to many problems in Geometry Processing and 3D
Shape Analysis problems, such as 
parameterization \cite{Kurtek2011},
shape retrieval \cite{Rabin2010},
isometry-invariant shape classification \cite{Bronstein2006},
mesh watermarking \cite{Bennour2006},
object recognition \cite{Hamza2003},
texture mapping \cite{Oliveira2011},
skinning \cite{Sloan2001},
just to mention a few. Because finding geodesic distances is a basic step in many geometrical
algorithms, the efficiency of its computation is an important issue.

Meshes are typically described as graphs. Consequently, one natural approach to solve the problem of
geodesic computation on meshes is to generalize the ideas of distances on graphs to compute distance
maps on surface representations. This path was pursued by Mitchell et al. \cite{Mitchell1987}, who
proposed a continuous Dijkstra method that is able to yield exact results. Later,
Surazhsky et al. \cite{Surazhsky2005} refined that method and proposed a more efficient approximate
version. Other researchers approached the problem via physical phenomena analogy. Two of the most
important methods in this class are based on models for wave propagation and heat diffusion. The
Fast Marching approach belongs to the first category of methods and aims to solve the so-called
Eikonal Equation \cite{Sethian1999}, \cite{Kimmel1998}. The Geodesics on Heat \cite{Crane2013}
belongs to the second category and explores the relationship between the heat kernel computation and distances on surfaces.

Here, we propose a parallel algorithm for the computation of distance maps on meshes that produces
results with competitive accuracy and is simple to implement. Like the Fast Marching method, our
method is also inspired by the grassfire propagation, but our approach does not require the
maintenance of any priority queue. Instead, we propagate distances simultaneously around the
frontier of propagation. This is the key to our parallelization strategy.

We provide an estimation of the complexity analysis (see section \ref{sec:analysis}) that justifies
the number of required iterations. Our experiments confirm that the estimation is not too tight, and
show that far less than $c\sqrt{n}$ iterations are necessary for the convergence of the method,
where $c$ is a small constant between $1 \leq c \leq 2$ and $n$ is the number of vertices of the
mesh. We also discuss and show that our method is especially appropriate for solving the
multi-source distance map problem (see \ref{ap:multi}).

\subsection{Contribution}

We propose a minimalist parallel method for computing geodesic distances on meshes with the
following properties: it does not require heavy pre-processing steps; it produces quite good accuracy
results even without pre-processing obtuse triangles; it is able to solve the problem for very large
meshes; it can be used with single precision configurations if necessary; and finally, it produces
speedup results comparable or better, in some cases, than the state-of-art methods.

We introduce an iterative parallel algorithm called \textit{\appname} (\appabrv) to compute
distance maps from a set of multiple sources on triangular meshes. Our method is inspired by the
Fast Marching algorithm and is based on the propagation of distance information from the inner to
outer groups of vertices that are equidistant to the source vertices. We call these groups
\textit{topological level sets} (\toplesets, for short). The vertices in each
\topleset have their distances updated in parallel, independently, enabling us to explore the
powerful parallel architectures present in the GPUs. To appropriately deal with coalescence problems
when accessing the vertices in the GPU memory we devised a breadth-first search based graph
representation similar to the one by Zhu et al. \cite{Zhu2017}.

Our method is particularly appropriate for distance computation from multiple sources when the
sources are predominantly distributed in a uniform way. This makes out method remarkably adequate
when used as part of the implementation of the Farthest Point Sampling algorithm \cite{Eldar1994}
for mesh resampling.

The proposed method is applied directly to the meshes and produces good accuracy results for
large and irregular meshes. Additionally, the speedup values are competitive against methods
where the preprocessing step dominates the overall complexity even without taking into account
the preprocessing time.

\subsection{Outline}

This paper is organized as follows. In Section \ref{sec:relatedwork}, we describe some of the most
important works related to distance computation on graphs and minimal geodesic computation on meshes.
Next, in Section \ref{sec:background}, we describe and define some concepts and results that were
used to develop our algorithm. The proposed method is presented in Section \ref{sec:approach}. In
Section \ref{sec:results}, we present the results produced by our algorithm. We first show two
applications of our method: a parallel solution to the Farthest Point Sampling Problem and a
parallel solution to the problem of computing the Voronoi Diagram on meshes. Next, we present the
speedup of our parallel method for different meshes and measure the distance error for each
experiment. We also compare our results with the exact method of \cite{Mitchell1987}, the
Fast Marching Algorithm of \cite{Kimmel1998} and the Geodesics on Heat method \cite{Crane2013}.
Finally in Section \ref{sec:conclusion} we present our conclusions.

\section{Related Work}\label{sec:relatedwork}

The exact computation of geodesic distances on surfaces was proposed by Mitchell et al.
\cite{Mitchell1987}. Their MMP algorithm is based on a continuous Dijkstra
algorithm and its computational complexity is $O(n^2 \log n)$, where $n$ is the number of
vertices. 

Due to this high computational cost, approximate methods were developed. These methods present
a better performance and maintain a comparable level of accuracy. A fast implementation of the MMP
algorithm was presented by Surazhsky et al. \cite{Surazhsky2005}; that algorithm requires less memory
and has a computational complexity of $O(n \log n)$. 

Over the last decade, several approximate methods were proposed to compute distances by solving the
Eikonal equation:
\begin{equation}
||\bigtriangledown \phi|| = 1 
\end{equation}
where $\phi$ is a distance function. We can divide these approaches into two families: the Fast 
Marching and the Fast Sweeping.

The Fast Marching method (FM) was introduced by Sethian to solve distance computation on regular
grids \cite{Sethian1999} and later extended to triangular meshes by Kimmel and Sethian
\cite{Kimmel1998}. The algorithm preserves the spirit of the Dijkstra algorithm
as it uses a priority queue. It is a single-source to all-vertices algorithm, whose main
advantage is the fast calculation of distances of vertices that are close to the source vertices.
However, due to the sequential requirement of the priority queue, it is not possible to parallelize
this algorithm without critical modifications. 

The Fast Sweeping approach \cite{Qian2007} has $O(n)$ linear computational complexity. However,
it requires a lot of sweeps to converge, particularly when the grids are unstructured.

A method based on Sethian’s Fast Marching method and Polthier’s straightest geodesics theory
\cite{Polthier1998} is proposed by Martinez et al. \cite{Martinez2005}. In this work, they propose
an iterative method to improve the discrete geodesic distances on triangulated meshes using inexact
algorithms. It starts computing an initial approximate discrete geodesic curve $\gamma_0$ using the
Fast Marching Algorithm. In the sequel, it applies a sequence of correction operation that computes
a sequence of discrete geodesic curves $\gamma_i$ by applying local correction operations. The local
correction operations aim at improving the positions of the discrete curve vertices and are inspired
by the definitions of \textit{straightest geodesics} and \textit{shortest geodesics} introduced by
Polthier and Schmies. The authors claim that their method can be used to improve the accuracy of
any inexact method that computes discrete geodesics.

A parallel version of the Fast Marching Algorithm on parametric surfaces was proposed by
Weber et al. \cite{Weber2008}. In this method, the surface must be divided into several regular
grids. Then, the distance map is computed for each grid, using a parallelization strategy based
on the Raster Scan algorithm \cite{Danielsson1980}. Finally, the reconstruction is done by joining
the distance maps associated with each grid via the Dijkstra algorithm. Up to now, this method is
the fastest and highly parallelizable for computing geodesic distances on triangular meshes. However,
the error of the computed distance map depends on the distortion of the parameterization.

A method that does not use a priority queue is the Fast Iterative method \cite{Fu2011}, which is an
algorithmic framework to solve the Eikonal equation. The main difference with our proposed method
is that they use an unordered list of vertices which are removed and added constantly until the
distances of the vertices converge, whereas our method keeps a structure of \toplesets where the
\toplesets that are removed cannot be added again to the set of vertices to be updated.

There exists approaches that do not solve the Eikonal equation, such as the Saddle Vertex Graph (SVG)
\cite{Ying2013}, which encodes the geodesic information in a sparse undirected graph, and the method
proposed by Wang \cite{Wang2017} which outperformed the SVG by using a divide and conquer technique.

Recently there was introduced a new family of methods that require the solution of Poison-like
systems. The Heat method was introduced by Crane \cite{Crane2013, Crane2017} and requires the
solution of the Poisson equation. This method can be used with different kinds of representations
because it is possible to compute the Laplacian operator for many different models, including
triangular meshes, point clouds and polygonal meshes. However, the accuracy of the distance map
computation is sensitive to the choice of a parameter. A parallel and scalable version of the Heat
method was proposed by Tao et al. \cite{Tao2018}. Finally, Litman and Bronstein \cite{Litman2016}
proposed a method that also works on the spectral domain, called the Spectrometer. Both methods
require as preprocessing the computation of the Laplacian matrix. Moreover, the Spectrometer
requires the computation of the Heat kernel, which involves computing the eigendecomposition and is
computationally expensive. In contrast, our method avoids heavy pre-processing steps.

\section{Background}\label{sec:background}

In the next subsections, we present some of the basic concepts and results on which our method is
based.

\subsection{Discrete distance maps}

Many graphical objects are described by 2-$d$ manifolds, embedded in a three-dimensional space,
that is, \textit{embedded surfaces}. Surfaces are usually represented by piecewise linear
representations known as \textit{triangle meshes} $T = (V,E,F)$, where $V$ is the set of
vertices, $E$ the set of edges and $F$ the set of faces. We now state the problem of computing
distance maps on a triangulated mesh $T$. 

Let $T$  be a triangulated mesh. Given a subset $S \subset V$, called source set, compute a map
$d_S(v):V \rightarrow \mathbb{R} $ that associates to each $v \in V$ its geodesic distance to $S$.
For the sake of simplicity, we denote $d_S$ as $d$.

Informally, we define \textit{geodesic distance} as the length of the minimum path connecting two
vertices of the mesh $T$ possibly passing through the faces of the mesh. In contrast, we use the
term \textit{topological distance} as the number of edges in the minimum path connecting two
vertices passing only through the meshes' edges.


\subsection{Fast Marching algorithm}

The Fast Marching algorithm (Algorithm \ref{algorithms/fastmarching}), simulates the propagation
of the distance information in a discrete set. It is possible to make an analogy with the
propagation of fire in a grassland or simply \textit{fire propagation}.
In the beginning, each source vertex $s \in S$ has its distance fixed to zero ($d(s) = 0, s \in S$)
and is inserted in a priority queue $R$ (red vertices), whose priority is defined in 
terms of the smallest distance. All other vertices $v \notin S$ are labeled with distance equal
to infinity ($d(v)=\infty$). At each step, one vertex $v$ is selected from the priority queue and
inserted in the list of processed vertices $B$ (black vertices); for all neighbor vertices
$v_0 \in \mathcal{N}(v)$ of $v$, all triangles incident to it $F(v_0)$ are updated using Algorithm
\ref{algorithms/planar_update}, which was proposed by Kimmel and Sethian in \cite{Kimmel1998}.
The new vertices on the updated triangles are inserted in the priority queue $R$ and the algorithm
proceeds until all vertices are processed, that is, $B = V$.

The update step (Algorithm \ref{algorithms/planar_update}) is one of the distinct features of the
Fast Marching method. It yields a linear local approximation to the continuous distance and 
guarantees that the solution obeys both the \textit{consistence condition} and the
\textit{monotonicity condition} ($QX^Tn < 0$) of the signal propagation.
When considered together, they ensure that, for a given triangle, defined by three vertices
$v_0,v_1,v_2$, if $v_1$ and $v_2$ are closer to the source set, then $v_0$ cannot be reached by
the signal before $v_1$ and $v_2$, producing a correct solution for the Eikonal Equation. In
geometrical terms, this means that the triangles in the mesh cannot be obtuse \cite{Bronstein2008}.
One solution for dealing with meshes that have such triangles is to subdivide them or to locally
unfold the mesh \cite{Kimmel1998}. For more details about the Fast Marching method, we refer to
the reader to the following references \cite{Kimmel1998} and \cite{Bronstein2008}.

\myalgorithm{algorithms/fastmarching}{Fast Marching (FM) \cite{Weber2008, Bronstein2008}}
\myalgorithm{algorithms/planar_update}{\textit{update\_step} \cite{Weber2008, Bronstein2008}}

Both exact and Fast Marching based methods rely, to a lesser or greater extent, on the use of
priority queues. This makes it hard to parallelize them. Thus, we
decided to completely abandon the use of priority queues in our proposed method. Instead of fixing
the final distance for the closest vertex, at each iteration, we update the distances on subsets of
vertices that are good candidates for the propagation of distance information. More precisely we
take advantage of the discrete topological structure of the mesh which can be decomposed in
topological distance level sets around the source vertices to propagate the information
simultaneously and independently in multiple phases until the distances converge.

\section{Proposed Method}\label{sec:approach}

The method proposed here works by simulating the distance information through an ordered set of
vertices using sequential iterations that refine the estimated distances. One of the main features
of our method is to exploit the natural ordering induced by the topology of the graph associated
with the mesh. Vertices with the same topological distance in the unweighted graph induced by
the mesh are grouped in sets defined here as \textit{topological level sets (\toplesets)}. The
\toplesets and their topological distances to the source are used to guide the propagation and
also mark off its scope. Our algorithm is an iterative algorithm that uses the preliminary ordering
defined by the topological distances as an initial estimate of the real continuous distance. At
each iteration, the distances of a subset of all topological sets, defining a band
(the \textit{update band}), are updated in parallel using a relaxation scheme, which is the update
function (Algorithm \ref{algorithms/planar_update}) of the Fast Marching method. This relaxation
process is applied for a number of iterations until all distances converge.

In this section, we define more precisely the concept of \textit{topological level sets (\toplesets)}
and describe the notion of \textit{topological level sets propagation}, which is the kernel of our
algorithm. Next, we describe the proposed algorithm and its parallel implementation on GPU. Finally,
we explain how to leverage the properties of our method to solve the multi-source version of the
problem.


\subsection{Topological level sets propagation}

The \textit{topological level sets propagation} relies on the concept of
\textit{topological level set (\topleset)}. 

\begin{definition}[\topleset $V_r$]
Let $G$ be the unweighted graph induced by the combinatorial topological structure of a triangle mesh
$T = (V, E, F)$ and $S\subset V$ the set of source vertices. A \topleset $V_r$ in $G$ is the set
of all vertices $v \in G$, such that the length of the shortest path from $v$ to the source vertices
$S$ in $G$ is equal to a constant $r \in \mathbb{N}$, $r>0$. A \topleset $V_r$ is defined by the
recurrence relation
$$V_r = \left\{v \in \mathcal{N}(V_{r - 1}) : v \not\in \bigcup_{r' = 0}^{r - 1} V_{r'} \right\}$$
where $V_0 = S$ and $\mathcal{N}(V_{r - 1})$ is the set of all the vertices that are neighbors of the
vertices $v \in V_{r-1}$.
\end{definition}

The properties below can be proved by inspection using the definition.

\begin{enumerate}
\item $\displaystyle\bigcup_{r = 0}^{\rho-1} V_r = V$
\item $\forall r,r' \in [0, \rho-1], r \not= r': \displaystyle V_r \cap V_{r'} = \emptyset $
\item $\displaystyle\sum_{r = 0}^{\rho-1} |V_r| = |V| $
\end{enumerate}
where $\rho$ is the number of \toplesets from $S \subset V$ in a triangular mesh.
Note that the \toplesets are 0-indexed because we start counting the \topleset $V_0=S$.

A \textit{topological level set propagation} is defined as the propagation of the distance
information through a sequence of \toplesets which defines a partial ordering on the vertices of
an unweighted graph $G$ induced by the mesh's combinatorial structure.




\subsection{An algorithm based on \topleset distance propagation}

\def\ptpband{B_{i_k}^{j_k}}


The proposed algorithm propagates and relaxes distance information through the meshes' \toplesets.
Algorithm \ref{algorithms/ptp} performs the distance propagation, updating at each
$k$-th iteration, the current distance map $d_k$, as a function of the previous distance map
$d_{k - 1}$ at iteration $k - 1$, for all vertices in the \textit{update band} $\ptpband$.

We define the \textit{update band} 
$$\ptpband = \displaystyle\bigcup_{r = i_k}^{j_k} V_r$$
for each iteration $0 < k \leq K$, where $K$ is the maximum number of iterations.
The \textit{update band} $\ptpband$ defines the set of vertices that will be updated in the current
iteration $k$, and is composed of a set of consecutive \toplesets $V_r$, $i_k \leq r \leq j_k$. The
whole set of consecutive \toplesets are computed with a breadth-first traversal in the induced graph.

We define the \textit{update band}'s boundaries $i_k$ and $j_k$ as sequences of values at each
iteration $k$ as follows:

\begin{equation}\label{eq:lower_band}
i_{k} = \left\{ 
	\begin{array}{ll}
	i_{k-1} + 1 & \text{if } \displaystyle\frac{\left|d_k(v) - d_{k-1}(v)\right|}{d_{k-1}(v)} < \epsilon, \forall v \in V_{i_{k-1}} \\
	i_{k-1}	& \text{otherwise} \\
	\end{array}
	\right.
\end{equation}

\begin{equation}\label{eq:upper_band}
j_{k} = \left\{ 
	\begin{array}{ll}
	k	& \text{if } k < \rho \\
	\rho - 1& \text{otherwise} \\
	\end{array}
	\right.
\end{equation}
where $\epsilon$ is a threshold for the relative change condition, $\rho$ is the \toplesets number, and
$d_k$ is the distance map at iteration $k$. This means that the lower boundary index is increased,
thus (shortening the band) if the relative change of the vertices in the corresponding \topleset is
smaller than an $\epsilon$ from one iteration to the next. In the experiments, we set
$\epsilon = 0.001$. Algorithm \ref{algorithms/ptp} (lines 12 to 18) computes at each iteration
the size of the \textit{update band}. A graphical example of this explanation is depicted in
Figure \ref{figures/update_band}. 

\myalgorithm{algorithms/ptp}{\appname (\appabrv).}

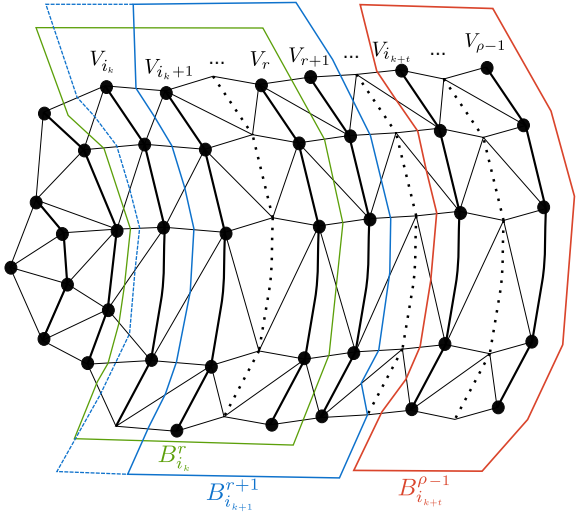
\image{figures/update_band}{0.9}{A graphical example of the \textit{update band} for the iterations $k$, $k + 1$ and some iteration $k+t \geq \rho$.}

One of the most important operations in the topological level sets propagation algorithm is the
update step (Algorithm \ref{algorithms/planar_update}), which is based on the Fast Marching method
update step. It is possible to update all vertices independently (line \ref{alg:ptp_iter_line},
Algorithm \ref{algorithms/ptp}), since each vertex $v$ updates its new distance $d_k(v)$ according
to the distance map $d_{k-1}$ computed in the preceding iteration $k-1$.

One distinct ingredient of the \appname (\appabrv) is that the \textit{update band} deals with the
fact that, as the frontier travels forward, there will be vertices whose distances have already
converged; vertices which are located in the first \toplesets of the propagation. Such vertices are
discarded by using the relative change based approach (lines 12 to 18, Algorithm
\ref{algorithms/ptp}) as mentioned before. 


\subsection{Parallelization and Implementation on GPU}

The \appname algorithm completely eliminates the dependency on the priority queue which is
necessary for the classical Fast Marching Algorithm. This permits us to reuse the calculations of
the previous distance maps to generate a new estimate. Furthermore, as the calculation of distances
is independent for each vertex $v_0$, the loop (see Line \ref{alg:ptp_iter_line}
in Algorithm \ref{algorithms/ptp}) is highly parallelizable on SIMD and GPU
processors.

Since meshes are irregular graphs, they induce random accesses to the global memory. Thus, the
non-coalescence problem needs to be handled carefully to avoid an inefficient implementation on GPU.
In this section, we explain how we deal with this problem. We were inspired by the method presented
in \cite{Zhu2017} which addressed the problem of iterative traversing-based graph processing, for
instance, Breadth First Search (BFS). Our method is quite similar to a BFS approach, with the
difference that traversing tasks are being performed at the same time. Thus, we adapted some ideas
from the WolfPath algorithm \cite{Zhu2017} to deal with meshes. To solve the problem of
non-coalescing memory access, they convert the graph into a layered tree representation using a
breadth-first traversal, where the vertices that are in the same layer are duplicated in the next
level. The data structure is a layered edge list composed of two arrays, a source, and a destination
array. Our method performs a similar step when it is used on a GPU. We compute the \toplesets and
we change the order of the data array that represents a mesh as a Compact Half Edge data structure
(CHE) \cite{Lage2005}. In this way, coalesced memory access is guaranteed.


\section{Experimental evaluation}\label{sec:results}

We have implemented two versions of the proposed algorithm using C++ with OpenMP and CUDA to execute
the experiments on CPU and GPU, respectively. We have made available the source code for the \appabrv
implementation as part of our framework \textit{gproshan (geometry processing and shape analysis
framework)} in \url{https://github.com/larc/gproshan}.

The experiments were performed in a Docker
container hosted in a machine with this configuration: an Intel Xeon E5-2698 v4 2.2 GH with 40
cores and 256 GB of RAM, a Tesla P100 with 16GB of RAM memory and 3584 cuda cores, and Ubuntu
16.04.3. The Docker container was set with the GCC/G++ 7.3 compiler, CUDA 9.0, and the required
libraries to run the algorithms. 

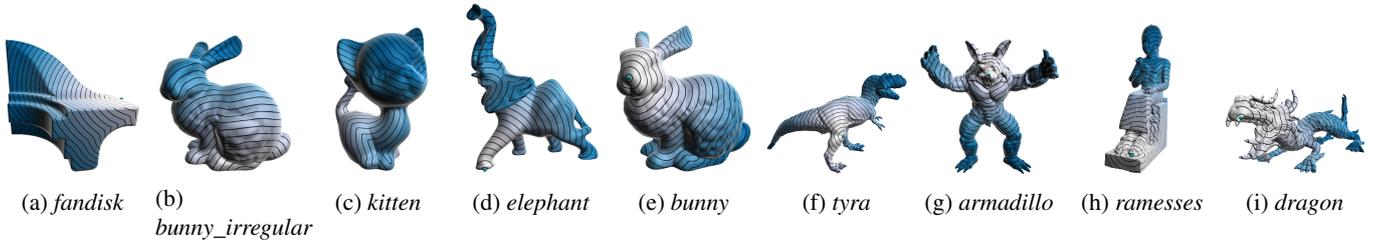
\mybimage{figures/ptp_meshes}{Geodesic distances map for each mesh in Table \ref{tables/ptp_tests}, from $m = 1$ sources.}
\mytable{tables/fm_tests}{FM from Toolbox Graph compiled with -O3 flag vs our FM implementation. Comparison time and mean absolute percent error (double precision). The experiment was performed on a Intel(R) Core(TM) i7-6700 CPU, 3.40GHz.}
\mybtable{tables/ptp_tests}{Single precision: comparison times, mean absolute percent error and speedup.}
\mybtable{tables/ptp_tests_double}{Double precision: comparison times, mean absolute percent error and speedup.}

We have implemented the Fast Marching algorithm that uses a priority queue, to compare its
performance with the performance of our algorithm. To validate our FM implementation, we compared its
performance with Gabriel Peyré's implementation used in the FM Toolbox Graph \cite{GabrielCode}.
Table \ref{tables/fm_tests} shows our test values. The time of our FM is better than Peyré's FM for
large meshes. For meshes, where the number of vertices are in the range of $25000$ and $100000$, we have competitive time.

Since our FM implementation is faster for large meshes, the speedup values obtained with the
\appabrv method are lower than what we would obtain with Peyré's implementation, so the time
comparison is reliable. We can observe that our FM's error is quite similar but not better than
Peyré's FM. In our implementations (our FM's code and \appabrv), we did not deal with the presence
of obtuse triangles by using unfolding or triangle subdivision. Nevertheless, as it can be seen
in the results, \appabrv produced quite good accuracy results even without pre-processing obtuse
triangles.

To compare the accuracy of the algorithms, we used the MMP algorithm proposed by Mitchell et al.
\cite{Mitchell1987}, to compute exact geodesics, which is included in the MeshLP package
\cite{MeshLP, GeodesicsCode}. We also compare our method with the Heat method proposed in
\cite{Crane2013, Crane2017}, which requires a heavy pre-processing step. However, after the
computation of the pre-processing step, the time complexity to calculate distance queries is
quite fast because the linear system is sparse. We chose this method because it is one of the
fastest in literature and its source code is available \cite{HeatCode}. 

We performed experiments on CPU and GPU, in single and double precision. The main advantage of
\textit{single precision} over \textit{double precision} is memory consumption. The
\textit{double precision} requires twice the number of bits than \textit{single precision} requires
to represent a decimal number.
The specific details of the implementation as well as the performance and accuracy results obtained
from the experiments will be given in the following sub-sections. The performance and accuracy are
evaluated over the meshes shown in Figure \ref{figures/ptp_meshes}. Tables \ref{tables/ptp_tests}
and \ref{tables/ptp_tests_double} summarize the experiments of distance computation from $m = 1$
source.
 
\subsection{Performance}


We evaluated the performance in single and double precision separately. The \appabrv method in
CPU was accelerated using the OpenMP library; the number of threads was 40. For the GPU
implementation, we used a block size of 64.

Table \ref{tables/ptp_tests} shows the execution times and speedup for the \appabrv algorithm
implemented on CPU and GPU with single precision. The speedup is considered over the FM algorithm.

The experiments with double precision are summarized in Table \ref{tables/ptp_tests_double}. We 
included the Heat method experiments only in this part because the solution of Cholesky
factorization requires double precision.
 
\image{figures/ptp_speedup_sources}{1}{\appabrv speedup with $m \in [1:1000]$.}

In Table \ref{tables/ptp_tests_double} we observe that the implementation of the Geodesics in Heat
method using the Cholmod library is fast and outperforms our method for small and medium
size instances using a hybrid GPU/CPU solver (using the Cholmod library). For larger instances, the
Geodesic in Heat method could not produce the correct results whereas the \appabrv computed the
correct distance maps with an approximate speedup of 55x for the \textit{dragon} mesh. Moreover, the
implementation of the Geodesics in Heat method using the Cholmod library requires the use of double
precision while the \appabrv can be run using both double and single precision. In the experiments,
we noticed that the Cholmod did not activate the GPU component using only the CPU cores. We
speculate that the reason for this is that the fill ratio of the Laplacian matrices of the meshes
is too small. The meshes we used in our experiments have a fill ratio between $5$ and $8$; we
compute these values according to the definition of fill ratio presented in \cite{Rennich2014}. We
can also claim that the \appabrv yields better accuracy results than the Geodesic on Heat for
irregular meshes like \textit{bunny\_irregular} and \textit{tyra}.

We can observe that in the single precision experiments the \appabrv method in CUDA outperforms
the \appabrv method on CPU. However, for small meshes like \textit{fandisk} and
\textit{bunny irregular}, both methods have similar speedup values. This behavior is expected
since the acceleration does not compensate for the GPU overhead for small meshes.

The Heat method requires the solution of a linear system. To this end, the usual implementation computes
the Cholesky factorization, which takes a considerable time to compute. However, once we calculate this
matrix, the time for the queries is quite fast. To accelerate the Heat method on the CPU, we used
Cholmod from SuiteSparse library \cite{Rennich2014}. However, we could not find a method that
works with sparse matrices and solves the problem entirely on the GPU device. We found the CUDA
library \textit{cusparse}, which is the fastest; the problem is that the conditions are incomplete
and therefore the results do not lead to a proper distance map. Another CUDA library we found is
the \textit{cusolverSp}, which uses a hybrid method as parts of the computation are done on
CPU and others on GPU; the use of \textit{cusolverSp}
produced the expected results, but it is still slow which can be explained by a possible overhead
due to multiple memory copies from the device to host and host to the device. Clearly, this is
a limitation of current CUDA libraries to implement the Heat method. It is difficult to get benefits
from powerful GPU devices when we use those implementations of the Heat method as is shown in Table
\ref{tables/ptp_tests_double}. The speedup values obtained from the Heat method with
\textit{cusolverSp} are quite low. Additionally, the solution to the Cholesky factorization is not
guaranteed, even with double precision. In our experiments, it was not possible to get a proper
solution for the largest meshes \textit{tyra}, \textit{armadillo}, \textit{ramesses} and
\textit{dragon}. In the case of CPU, the Cholmod library is more stable and we got a good result
for the \textit{armadillo} mesh. Recently, Tao et al. \cite{Tao2018} proposed a scalable version of
the Heat Method, which has been tested only on CPU for large meshes. This method uses Gauss--Seidel
iterations instead of solving the linear system using Cholesky decomposition. The method also
improved the memory consumption of the previous implementation of the Heat method. However, it
still depends on the choice of a good parameter, which for irregular meshes still presents a
higher error on the distance computation.

In contrast, the \appabrv method successfully computes the distance map using single and double
precision. Moreover, as expected, the speedup values obtained for the \appabrv method, using
single precision are faster than the speedup values obtained using double precision.

In Section \ref{sec:analysis}, we showed the importance of the inverse relationship between the
complexity of the \appabrv algorithm and the number of sources and the way they are distributed. We
also discussed how the distance can be computed independently from each source at each iteration
allowing the parallel implementation.

Figure \ref{figures/ptp_speedup_sources} presents the speedup of the \appabrv algorithm for each 
mesh from $m$ sources with approximate uniform distribution. For this experiment we did not 
perform the rearrangement of the vertices on the GPU memory, since there are multiple sources.
Note that the speedup of meshes with a
large number of vertices increases with the number of sources $m$; this illustrates how our 
algorithm, performs well when computing distances from multiple
sources and also reinforces the expected inverse relationship between performance and number of
sources in the proposed \appabrv algorithm.

\subsection{Accuracy}\label{subsec:accuracy}
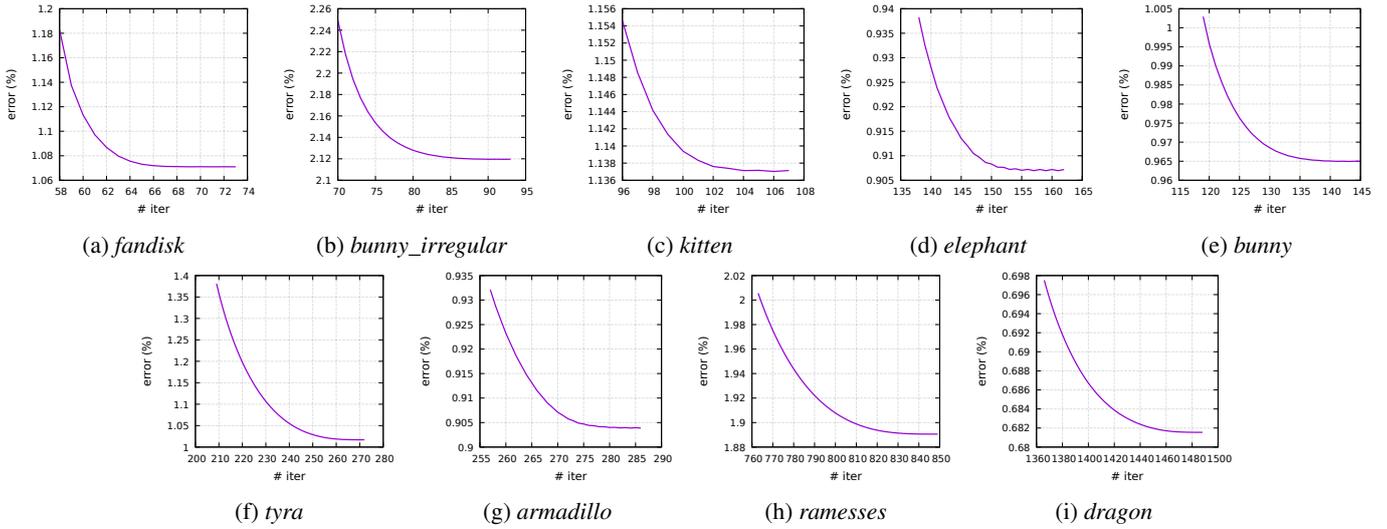
\mybimage{figures/ptp_iter_error_double}{Mean absolute percent error per number of iterations.}

We tested the accuracy of the proposed algorithm, performing a comparison based on the
Mean Absolute Percent Error (MAPE) between the FM algorithm and the \appabrv algorithm,
summarized in Tables \ref{tables/ptp_tests} and \ref{tables/ptp_tests_double}. We can observe
that the \appabrv algorithm has error (MAPE) values similar to the FM algorithm, and even less for
\textit{fandisk} and \textit{dragon} meshes. We can improve \appabrv's accuracy, modifying
the $\epsilon$ value for the relative error. This happens because we neither performed mesh
unfolding nor subdivided obtuse triangles, which are necessary steps for the FM algorithm. We show
experimentally that our algorithm corrects the Dijkstra's estimate in the case of obtuse triangles
in subsequent iterations.

We evaluated the convergence of the \appabrv algorithm as a function of the number of iterations.
The experimental results show that no more than $c\sqrt{n}$ iterations are required for the \appabrv
convergence, with $c \approx 1.5$ for predominantly regular meshes An analysis of such behavior is
discussed in Section \ref{sec:analysis}. 

\simages{0.4}{{
figures/circle_rp/,
figures/circle_fm/,
figures/bunny_rp/,
figures/bunny_fm/
}}
{fig:rp_fm}{Level sets propagation in \ref{figures/circle_rp} and 
\ref{figures/bunny_rp},
geodesics distances map in \ref{figures/circle_fm} and \ref{figures/bunny_fm}.}

Figure \ref{figures/ptp_iter_error_double} presents the empirical absolute mean errors as a function
of the number of iterations; each chart shows values starting with an iteration $k = \rho$
(number of \toplesets of the mesh) and finishing within $K$ (number of iterations) empirically
estimated by our algorithm for each mesh. Note that the error decreases as the number of iterations
are increased and it converges for some $k$ between $\rho$ and $K$. The exact number of iterations
$K$ depends on the topology of the triangular mesh and the geometric relation between the propagation
of level sets and the geodesic distance map. 

The relationship between the level sets propagation and the geodesic distance map is depicted in
Figure \ref{fig:rp_fm}. The topological level sets propagation and the isocurves defined by the
distance map are quite similar on the disk. This means that the geodesic curves are almost
perpendicular to the discrete curves defined by the \toplesets and that the initial edge-distance is
a good approximation to the final continuous distance. In this case, our algorithm only needs about
$\rho$ iterations to compute the correct distance map. For the bunny mesh, we observe a similar
behavior. Nevertheless, as the \topleset arrangement diverges more from the final isocurves of
distance, we need more than $\rho$ iterations, but no more than $1.5\rho$ as expected for
well-behaved meshes; see Figure \ref{figures/ptp_iter_error_double}.


\subsubsection{Comparison with the Heat method}

One of the main problems with the Heat method was the choice of the parameter. According to
the paper \cite{Crane2017}, a good approximation is the square mean of the mesh edges' length. However in case of
the \textit{ramesses} mesh it was not possible to obtain a correct result. For this reason, in
Table \ref{tables/ptp_tests_double}, the error is $nan$. Another limitation of the considered
implementation of the Heat method is that the use of double precision is mandatory since this
implementation of the Heat method requires to solve the Cholesky factorization. 
Furthermore, even with double precision,
the CusolverSparse library does not always produce a solution when dealing with large meshes.
In our experiments, it crashed for the \textit{armadillo} and \textit{ramesses} meshes. So, we got
an \textit{inf} error. This is another limitation of this implementation of Heat method
because it depends on the Cholesky factorization for sparse matrices and this computation is still
unstable, especially in libraries for GPU devices.

In contrast, our method works faster
when dealing with single precision and the accuracy is little affected. These results are shown in
Tables \ref{tables/ptp_tests} and \ref{tables/ptp_tests_double}.

Finally, in Table \ref{tables/ptp_tests_double}, we note that for the \textit{irregular bunny}, the
Heat method presents a higher error than the proposed method. Thus, our method is robust when
dealing with irregular meshes and obtuse triangles because it is able to correct the error through
multiple iterations.

To summarize, we showed how our method scales, has competitive speedup values and yields better 
accuracy results.

\subsubsection{\appabrv performance on meshes with obtuse triangles}

In the next experiments, we show that the presence of obtuse triangles affects the accuracy of the
\appabrv and FM methods. We compared the accuracy of the methods in a regular grid mesh, shown in
Figures \ref{figures/grid_regular} (input grid), \ref{figures/grid_regular_fm} and
\ref{figures/grid_regular_gpu}, that shows the distance maps for FM and \appabrv methods,
respectively. Both algorithms have the same error of $0.1229\%$. Though, under the presence of
obtuse triangles, the precision of the methods is slightly different.

\simages{0.3}{{
figures/grid_regular/$1002001$ vertices,
figures/grid_regular_fm/FM: $0.1229\%$,
figures/grid_regular_gpu/\appabrv: $0.1229\%$,
figures/grid_obtuse/$1170$ vertices,
figures/grid_obtuse_fm/FM: $3.754\%$,
figures/grid_obtuse_gpu/\appabrv: $2.799\%$
}}
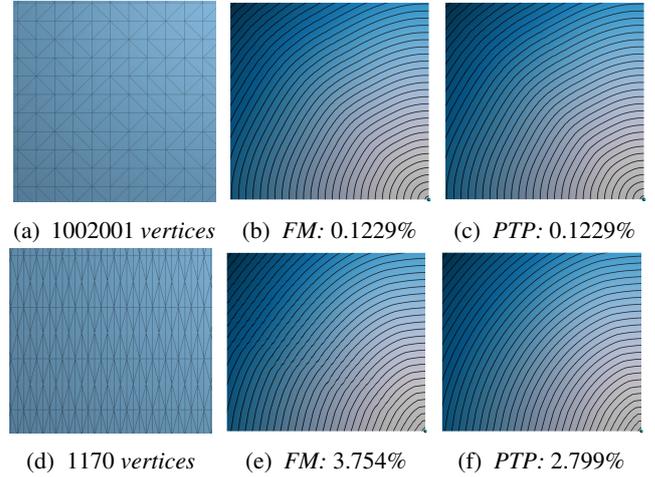
{fig:obtuse_test}{Distance maps and error comparison between a regular triangular mesh and an obtuse triangular mesh.
Figure \ref{figures/grid_regular} shows a piece of a regular mesh,
\ref{figures/grid_regular_fm} shows the FM's distance map computed for \ref{figures/grid_regular},
and \ref{figures/grid_regular_gpu} shows \appabrv's distance map computed for \ref{figures/grid_regular}.
Figure \ref{figures/grid_obtuse} shows a piece of an obtuse triangular mesh,
\ref{figures/grid_obtuse_fm} shows the FM's distance map computed for \ref{figures/grid_obtuse},
and \ref{figures/grid_obtuse_gpu} shows \appabrv's distance map computed for \ref{figures/grid_obtuse}.
}

We created an artificial
mesh which contains obtuse triangles, to measure the robustness of the proposed method against
the FM method. We are aware that FM method requires a pre-processing step, called unfolding, to
remove all the obtuse triangles of the mesh. However, we omitted this step in both the \appabrv
and FM to see how they behave in such conditions. The initial obtuse mesh is depicted in Figure
\ref{figures/grid_obtuse}. This is a mesh, which contains $1170$ vertices. The distance map
obtained with the FM is depicted in Figure \ref{figures/grid_obtuse_fm}. The result produced by the
\appabrv method is shown in Figure \ref{figures/grid_obtuse_gpu}. We observe that the distance map
produced by the \appabrv method is smoother and presents less error $(2.799\%)$ than the FM method 
$(3.754\%)$. We believe that this outcome is due to the distance correction or improvement that our
algorithm performs in each iteration. The greedy behavior of Fast Marching in these cases is not
suitable because obtuse triangles introduce error, which increases as the propagation moves away
from the origin. Thus, we believe, according to the experiments, that our method is more robust
than the FM algorithm when dealing with obtuse triangles without unfolding or subdivision.

\subsection{Farthest Point Sampling (FPS) and Voronoi Diagrams}

The Farthest Point Sampling (FPS) is a generic algorithm introduced by Eldar \cite{Eldar1994} that
generates a regular sampling. At each iteration, the farthest vertex to the current set of samples
$S$ is computed and inserted into $S$.

The FPS algorithm exploits the fact that, in the \appabrv algorithm, the performance increases when
the number of samples grows. This happens because the FPS algorithm in each iteration computes a
new sample, which is added to the set of sources $S$. This case is very favorable to the \appabrv
algorithm as the FPS tends to generate uniformly distributed samples.

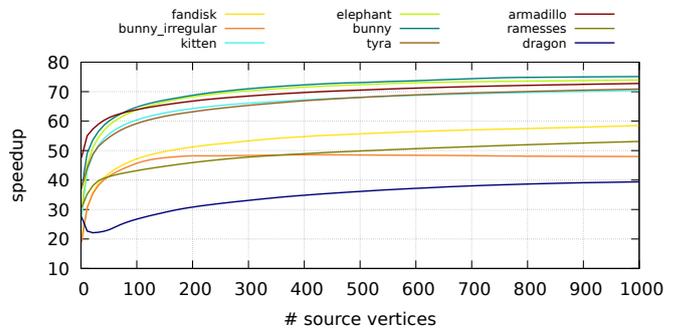
\image{figures/fps_speedup_sources}{1}{Farthest Point Sampling speedup with $m \in [1:1000]$.}

\simages{0.32}{{
figures/bunny_voronoi/,
figures/armadillo_voronoi/,
figures/ramesses_voronoi/
}}
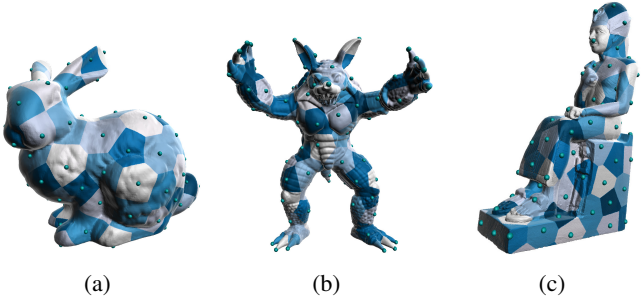
{fig:fps_voronoi}{Computation of the Voronoi
regions in \ref{figures/bunny_voronoi}, \ref{figures/armadillo_voronoi} and
\ref{figures/ramesses_voronoi}; from the calculation of $m = 100$ samples with the FPS algorithm.}

Figure \ref{figures/fps_speedup_sources} presents the speedup of the computation of $m$ samples,
using the FPS algorithm based on the GPU implementation of the \appabrv algorithm to compute the
geodesic distances. Figure \ref{fig:fps_voronoi} presents some visual
results for the computation of $m$ samples using the FPS algorithm with the geodesics distances
computed with the \appabrv algorithm; it also presents the Voronoi regions with $m$ sources, which
were computed with the FPS algorithm.

\section{Estimative complexity analysis}\label{sec:analysis}

In this section, we provide an estimative complexity analysis for our algorithm. To this end,
instead of considering an update band based on the relative change, as in Algorithm
\ref{algorithms/ptp}, we use a fixed update band. This modification enables us to define an upper
bound to our algorithm complexity. We can classify the \toplesets in monotonically increasing,
stationary and monotonically decreasing sequences. These \toplesets sequences are present in any
triangular mesh. Figure \ref{figures/toplesets_meshes} shows the number of vertices per \topleset
for each mesh in the experiments. We can observe that for each mesh's \toplesets, the first ones
shape a monotonically increasing sequence, the last \toplesets shape a monotonically decreasing
sequence, and the \toplesets in the middle can alter among monotonically increasing, monotonically
decreasing and stationary sequences.

\mybimage{figures/toplesets_meshes}{Number of vertices per \topleset for each mesh in the experiments.}

A finite sequence of topological level sets $(V_i)_{i=a}^b$, is a set of contiguous topological
level sets where $a$ is the starting index of the sequence and $b$ is the ending index. We classify
a sequence $(V_i)_{i=a}^b$ according to the behavior of the growth of the cardinality of each
\topleset that belongs to it, as follows:

\begin{enumerate}
\item \textit{Monotonically increasing \topleset sequence}:
 when $|V_i|-|V_{i-1}|>0$, $a \leq i \leq b$
\item \textit{Stationary \topleset sequence}:
 when $|V_i|-|V_{i-1}|=0$, $a \leq i \leq b$
\item \textit{Monotonically decreasing \topleset sequence}:
 when $|V_i|-|V_{i-1}|<0$, $a \leq i \leq b$
\end{enumerate}

Now, let $T$ be a triangle mesh with $\rho$ \toplesets and $n = |V|$. $T$ is a
\textit{monotonically increasing mesh} when all of its \toplesets define one monotonically
increasing sequence $(V_i)_{i=0}^\rho$. Similarly, $T$ is a \textit{monotonically decreasing mesh}
when all of its \toplesets define one monotonically decreasing sequence $(V_i)_{i=0}^\rho$. When
all its \toplesets have the same cardinality, except for a small finite number of \toplesets $c<<n$,
then the set of all \toplesets form a stationary sequence $(V_i)_{i=c}^\rho$ and we say that $T$ is
a \textit{stationary mesh}. In any other case we call $T$ an \textit{arbitrary mesh}.

The proposed algorithm propagates and relaxes distance information through the meshes' \toplesets.

At each iteration $0<k<K$, where $K$ is the maximum number of iterations, we define a band $B(k)$,
called \textit{update band}, that defines the vertices that will have their distances updated.
$B(k)$ is composed of a set of consecutive \toplesets $V_i$, $s(B(k)) \leq i \leq k$,
where $s(B(k))$ is the index of the first \topleset in $B(k)$. $B(k)$ and $s(B(k))$ are computed
using a breadth-first traversal through all consecutive \toplesets $V_i$
(see Section \ref{sec:complexity}).

\subsection{Complexity analysis sketch and comparison}\label{sec:complexity}

We begin the complexity analysis of the \appname algorithm without taking the number of threads
$\mathcal{T}$ into consideration. In the final part of the analysis, we discuss the impact of
$\mathcal{T}$ which will have a strong effect in the throughput when $\mathcal{T} \approx \sqrt{n}$.

The first step in the algorithm is to define the order used to propagate the distances through the
topological level sets. This can be done in linear time if the mesh is structured using a half-edge
or any similar topological data structure. Now we discuss the definition of the size of the band
$B(k)$ at each iteration $k$. The size of the band depends on the combinatorial structure of the
mesh but can be defined for each iteration k corresponding to each \topleset in monotonically
increasing, decreasing and stationary subsequences of the mesh. This step requires segmenting the
mesh's \toplesets in such subsequences. To obtain such segmentation, it suffices to start with a
given \topleset $V_0$ and analyze the behavior of the function $2(H[|V_1|-|V_0|] - 1/2) $ where
$H[n]$ is the Heaviside step function. To classify the first sequence into increasing $(+1)$,
decreasing $(-1)$ or stationary $(0)$, we iterate through the \toplesets using a breadth-first
traversal until $2(H[|V_1|-|V_0|] - 1/2)$ changes, signalizing the beginning of a new subsequence.
The process continues until all \toplesets are grouped into monotonically increasing, decreasing
and stationary subsequences. We show later in the text that for monotonically increasing subsequences
the size of the band is linear and equal to $k/2$, where $k$ is the order of the iteration, and for
monotonically decreasing and stationary subsequences the size is equal to one.

The next step that the \appname (Algorithm \ref{algorithms/ptp}) executes is
to relax the vertices' distances using the update function for a total of $K$ iterations. 
Consequently, the complexity of the \appabrv algorithm is given by the total number of update
operations $f(n)$, in the main loop of the Algorithm \ref{algorithms/ptp},
expressed as a function of the number of its vertices $n = |V|$ :

\begin{equation}\label{eq:n_operations}
f(n) = c\displaystyle\sum_{k=1}^K \sum_{r = s(B(k)) }^k |V_r|
\end{equation}
where $K$ is the number of iterations, $c$ is a constant, and $s(B(k))$ is the index of the first
topleset in $B(k)$.

Let $V_{r'} = \displaystyle\max_{r \in [0:\rho-1]} |V_r|$ be the topological level set with the
largest number of vertices, and $|V_{r'}|$ the maximum number of vertices in any \topleset. Thus, we claim
that:

$$c\displaystyle \sum_{k=1}^K \sum_{r = s(B(k)) }^k |V_r|
\leq c\displaystyle\sum_{k=1}^K \sum_{r = s(B(k)) }^k |V_{r'}| $$

$$f(K) \leq c |V_{r'}| \displaystyle\sum_{k=1}^K \sum_{r =s(B(k)) }^k 1 $$

Here, $s(B(k))$ denotes the index of the starting \topleset of the band $B(k)$. 
The number of iterations $K$, the number of vertices $|V_{r}|$ in a \topleset $V_r$ and the depth of
the bands $|B(k)|$ all depend on the combinatorial topology of the mesh and are intricately coupled
with each other. Nevertheless, an analysis can be done for meshes with specific \topleset cardinality
growth.
 
Let us now analyze the complexity of each one of the cases and posteriorly express the worst case
complexity for an arbitrary mesh. 

\subsubsection{Stationary mesh case}

Let the complexity of a \textit{stationary} mesh be defined as a function $g(n)$. For a stationary
mesh, the number of \toplesets multiplied by its constant cardinality $|V_i| = |V_{r'}|$ is
$\rho.|V_{r'}| = O(n)$. To define the size of $B(k)$ for a \textit{stationary} mesh we use the result
from the following lemma, which is proved as a special case of Lemma \ref{lemma:max_n_iter} in
the \ref{proof:max_n_iter}.

\begin{restatable}{lemma}{lemmanumiterconstmesh}
[Number of iterations necessary to fix the distances of a \textit{topleset} of a stationary mesh]
\label{theorem:max_n_toplesets}
Given a stationary mesh $T$, and a \topleset $V_k$, the number of iterations necessary to fix the
distances of all vertices $v \in V_k$ is equal to $1$.
\end{restatable}

Thus, the size of the band used in all iterations of a stationary mesh is equal to $1$. As 
each \topleset requires one iteration to fix all its distances and the total number of \toplesets is
$\rho$, then we claim that $K = \rho$. Finally, we state that the complexity of the proposed
algorithm when applied to a stationary mesh is:

$$g(n) \leq c\displaystyle |V_{r'}| \sum_{k=1}^K (k-s(B(k)))
= \displaystyle\ c|V_{r'}| \sum_{k=1}^K 1
= \displaystyle\ c|V_{r'}| K $$

$$g(n)= O(n)$$

\subsubsection{\textit{Monotonically increasing} mesh case}

Let the complexity of a \textit{monotonically increasing} mesh be defined as a function $h_1(n)$. For
a \textit{monotonically increasing} mesh, Theorem \ref{theorem:max_n_toplesets} claims that
$\rho = O(\sqrt{n})$.

\begin{restatable}{theorem}{theoremmaxntoplesets}
[Number of \toplesets $\rho$ of a monotonically increasing mesh]
\label{theorem:max_n_toplesets}
Let $T = (V, E, F)$ be a triangular mesh and $s \in V$ be a source vertex such that the cardinalities
of its \toplesets, ordered according to their distances to the sources, define a monotonically
increasing sequence. Then the number of \toplesets $\rho$ from $s$ satisfies
$\rho = O\left(\sqrt[\,]{n}\right)$, where $n = |V|$.
\end{restatable}

A proof to Theorem \ref{theorem:max_n_toplesets} is given in the \ref{proof:max_n_toplesets}.
In the same way, we can prove that for a triangular mesh
$T=(V, E, F)$ and a set of uniformly distributed source vertices $S \subset V$, the number of
\toplesets is $\rho = O\left(\sqrt[\,]{\frac{n}{m}}\right)$, where $m = |S|$.

As the number of \toplesets for a monotonically increasing mesh is $\rho = O\left(\sqrt[\,]{n}\right)$
and we are limited to $n = |V|$, the largest possible \topleset has cardinality equal to
$|V_{r'}| = O\left(\sqrt[\,]{n}\right)$.

To define the size of the band $B(k)$ for a monotonically increasing mesh, we rely on the partial
result presented in the proof in \ref{proof:max_n_iter}. It defines the number of iterations $K_r$
necessary to update all vertices in a \topleset $V_r$ as 

\begin{equation}\label{eq:n_operations}
K_r \leq \left\lceil\frac{(\Delta v-3)}{2}\right\rceil(r-1)+1
\end{equation}
where $\Delta v$ is the largest degree of the vertices $v \in V_r$. For a regular monotonically
increasing mesh, whose valency is $6$, we have $K_r = 2r - 1$; we can observe that in the
$K_r$-iteration the vertices in the \topleset $V_r$ with order $r = \frac{K_r + 1}{2}$ have their final distances
computed; then we can define an updating band between all vertices included in the level sets
$r = \left[\left\lfloor \frac{K_r + 1}{2} \right\rfloor, K_r\right]$.

Moreover, Theorem \ref{theorem:max_n_iter} states that for monotonically increasing meshes the
maximum number of iterations is $K = O\left(\sqrt[\,]{n}\right)$.

\begin{restatable}{theorem}{theoremmaxniter}
[Maximum number of iterations $K$ to compute the distances of a monotonically increasing mesh]
\label{theorem:max_n_iter}
Let $T = (V, E, F)$ be a triangular mesh and $s \in V$ be a source vertex such that the set of its
toplesets $V_r$ ordered by their distances to $s$ define a monotonically increasing sequence. Then,
the maximum number of iterations is $ K = O(\sqrt{n}) $, where $n = |V|$.
\end{restatable}

A sketch of the proof of Theorem \ref{theorem:max_n_iter} is given in the
\ref{proof:max_n_iter}, which is supported by the experimental
evaluation in Section \ref{sec:results}. Based on the Theorem \ref{theorem:max_n_iter},
we can compute an upper bound for the function $h_1(n)$ which defines the complexity behavior of the
\appabrv for monotonically increasing meshes.

$$h_1(n) \leq c\displaystyle |V_{r'}| \sum_{k=1}^K (k-s(B(k)))$$
$$h_1(n) = \displaystyle\frac{c}{2} |V_{r'}| \sum_{k=1}^K k = \displaystyle\frac{c}{4} |V_{r'}| K(K + 1) $$

$$h_1(n) = O\left(\displaystyle |V_{r'}| (K^2 + K) \right) = O\left(\displaystyle |V_{r'}| K^2 \right)$$
$$h_1(n) = O\left(\displaystyle |V_{r'}| {\sqrt{n}}^2 \right) = O\left(\displaystyle n\sqrt{n} \right)$$

\subsubsection{Monotonically decreasing mesh case}

The complexity of a monotonically decreasing mesh is expressed here as a function $h_2(n)$.
Similarly to the previous case, the upper bound to the number of \toplesets for a monotonically
decreasing mesh is $\rho = O(\sqrt{n})$. Although we do not present a formal proof for this case
here, this property can be verified considering that this case is symmetrical to the case of
\textit{monotonically increasing} meshes.

Differently, though, the number of iterations necessary to fix all vertices in a \topleset $V_r$ in
a monotonically decreasing mesh is similar to the stationary case and equal to $K_r=1$. This occurs
because the number of vertices from \topleset $V_{r-1}$ to $V_r$ decreases and the edges in the chain
associated to \topleset $V_{r-1}$ contains a sufficient number of vertices with fixed distances to
set the correct distances of all vertices in $V_r$. Consequently, $|B(k)|=1$ and the number of
iterations is $K=\sqrt{n}$. Moreover, in this case $|V_r| = O(\sqrt{n})$

Hence, we claim that:

$$h_2(n) \leq c\displaystyle |V_{r'}| \sum_{k=1}^K (k-s(B(k)))
= \displaystyle c |V_{r'}| \sum_{k=1}^K 1$$
$$h_2(n) = \displaystyle c|V_{r'}| K = \displaystyle c' \sqrt{n}\sqrt{n} = O(n)$$

\subsubsection{General case}

An arbitrary mesh is composed of a combination of segments that can be
\textit{monotonically increasing}, \textit{monotonically decreasing} and \textit{stationary}. Let $T$
be a mesh such that its \toplesets $V_i$, $0 \leq i \leq \rho$ are grouped into contiguous
\textit{monotonically increasing}, \textit{monotonically decreasing} and \textit{stationary}
sequences $s_j = (V_i), k_1(j) \leq i \leq k_2(j)$ where $0 \leq k_1(j), k_2(j) < \rho$. Let us
denote by $|s_j| = \sum_{V_i \in s_j} |V_i|$ the amount of vertices in all \toplesets of a sequence
$s_j$. Let $I$, $D$ and $C$ denote, respectively, the sets of \textit{monotonically increasing}
($si$), \textit{monotonically decreasing} ($sd$) and \textit{stationary} segments ($ss$) of
toplesets. The sum of all vertices in $I \cup D \cup C$ is equal to $n = |V|$. The number of update
operations in all segments is given by:

$$f(n) = \sum_{si_i \in I} g(|si_i|) + \sum_{sd_i \in D} h_1(|sd_i|) + \sum_{ss_i \in C} h_2(|ss_i|)$$
$$f(n) = \sum_{si_i \in I} |si_i| + \sum_{sd_i \in D} |sd_i| \sqrt{|sd_i|} + \sum_{ss_i \in C} |ss_i|$$
$$f(n) \leq c_1 n + c_2 n \sqrt{n} + c_3 n$$
$$f(n) = O\left(n\sqrt{n}\right)$$

\subsubsection{Multi-source problem}\label{ap:multi}

The complexity of the proposed algorithm is impacted not only by the size of the mesh but also by the
number $m$ of source vertices $s \in S$. For special cases, in particular, when the source
vertices are uniformly distributed, the number of \toplesets can be shrunk by a factor of $m$.

For arbitrary meshes, the complexity is dominated by the set of monotonically increasing subsequences
which depend on the number of vertices $n$ and the diameter of the mesh $\sqrt{n}$. Hence when the
number of \toplesets shrinks to $\sqrt{\frac{n}{m}}$ we have an overall complexity of:
$O\left(\frac{n\sqrt{n}}{\sqrt{m}}\right)$. 

This will have a great impact on the use of the \appabrv algorithm for implementing the
\textit{Farthest Point Sampling Algorithm}, an algorithm used to sample a mesh uniformly which 
will be described in \ref{app:algfps}.

\subsubsection{Complexity of the parallel implementation}

For a parallel implementation of the \appabrv algorithm with a number $\mathcal{T}$ of threads, applied
to an input given by $m = |S|$ sources and a mesh with $n = |V|$ vertices and $\rho$ \toplesets, the 
final worst case complexity is $O\left(\frac{n\sqrt{n}}{\mathcal{T}\sqrt{m}}\right)$. 


\subsubsection{Application: Farthest Point Sampling}\label{app:algfps}



The computational complexity analysis of FPS is $O(mn\log n)$ using FM, whereas the parallel version
of FPS using the proposed \appabrv has complexity of $O(\sqrt{m}n^{3/2})$. We present this
complexity analysis in Section \ref{app:fps}. The experimental results demonstrate that as
the number of samples grows, there is an increase in the performance of the algorithm.

\mytable{tables/complexity}{Comparison of complexity algorithm analysis.}

Table \ref{tables/complexity} summarizes the analyzed algorithms. We include the term $\mathcal{T}$
to the proposed algorithm because the parallelization is a feature in our algorithm, and also has
impact in the FPS algorithm.

\subsubsection{Note}

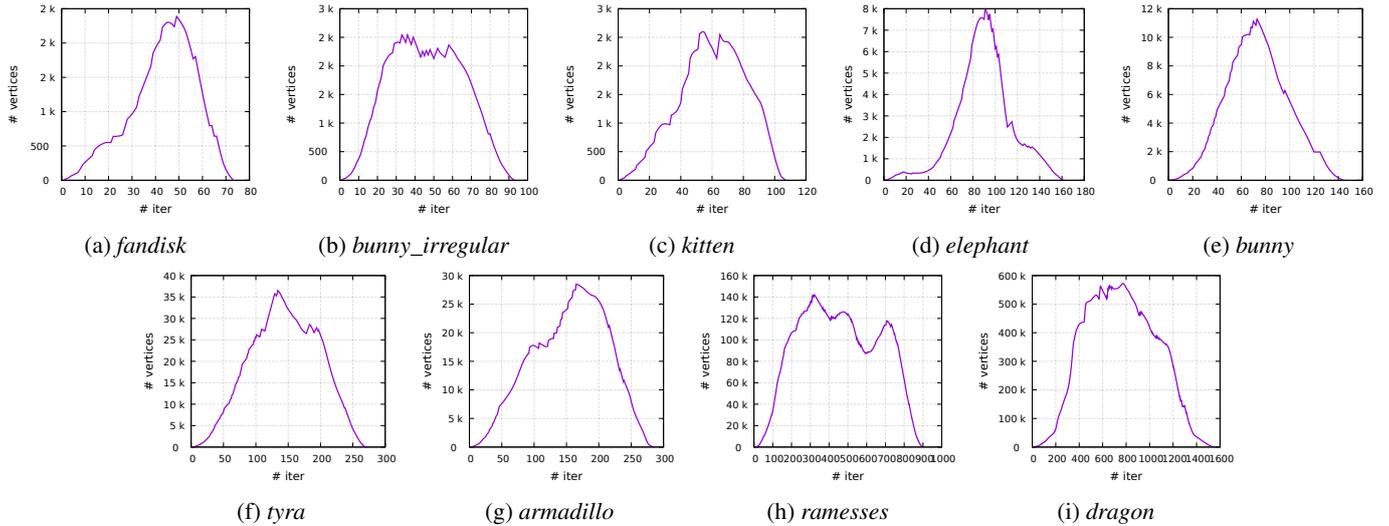
\mybimage{figures/ptp_band_vertices}{Number of updated vertices per iteration.}
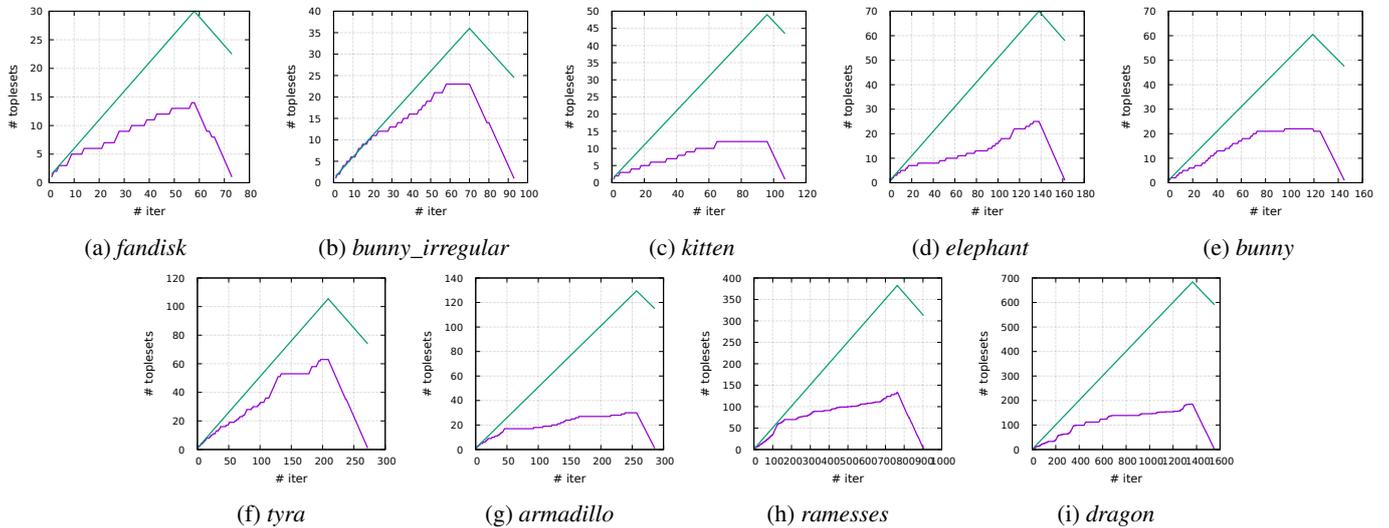
\mybimage{figures/ptp_band_toplesets}{Comparing the number of \toplesets per iteration between the dynamic band (purple curve), from empirical result for the Algorithm \ref{algorithms/ptp}; and fixed band (green curve) with size equal to $k/2$ \toplesets, considered in the analysis and sketch proofs.}


While the current complexity analysis, based on the sketch proof, states that the complexity of
our algorithm has an upper bound of $O\left(n \sqrt{n} \right)$, it is not tight enough if we
consider the behavior of our algorithm for regular and non-regular meshes as it is shown in the
experiments. The number of iterations $K$ necessary for our algorithm to converge is bounded by the
number  of \toplesets $ \rho $; it is $ K \leq c \rho $. The experimental results depicted in
Figure \ref{figures/ptp_iter_error_double}, show that $c \approx 1.5 $.

Figure \ref{figures/ptp_band_vertices} shows the total number of vertices that are updated in
each iteration of our algorithm and Figure \ref{figures/ptp_band_toplesets} depicts the comparison
of the number of \toplesets per iteration between the fixed band (green curve) and the dynamic band
(purple curve). The dynamic band is purely empirical, because it depends on the relative change
of the previous iteration (\textit{update band} in Algorithm \ref{algorithms/ptp}). The fixed band
set the number of \toplesets to be updated per iteration according the analysis of the algorithm
asymptotic complexity. We can observe that the green curve requires more iterations to converge
and also it updates more \toplesets per iteration.
In contrast, the purple curve shows that the algorithm updates a smaller number of \toplesets per
iteration and also it requires fewer iterations to converge. Note that in the case of irregular
meshes like \textit{bunny\_irregular} and \textit{tyra}, the empirical curve is closer to the
fixed curve. We left as future work, finding a tighter upper bound for the algorithm asymptotic
complexity.

\section{Conclusion}\label{sec:conclusion}

We have presented a minimalistic parallel algorithm to compute approximate distance maps on
triangular meshes. We also provide an implementation of our proposal on GPU. 

The main advantages of our approach are:
\begin{itemize}
\item {\bf Minimalistic:} We believe our method fits the class of minimalistic methods
because it uses the fewest elements as possible to achieve the best results. 
The method itself takes into account the topological structure (\toplesets) of the mesh and
take advantage of it to correct the distances using only the necessary iterations. At the same time,
the nature of our algorithm enables us to leverage powerful parallel architectures. 
Besides, our method avoids complicated pre-processing steps and dependence on parameters
that are difficult to set up. 

\item {\bf Scalable:} The distances are computed simultaneously through multiple
iterations, instead of using a priority queue. We have demonstrated how our algorithm scales
and leverages the use of GPU devices. Furthermore, the proposed method improves the speedup
considerably when used to solve the multi-source problem.

\item {\bf Memory consumption management:} The proposed method supports single precision and
does not require double precision to achieve a good accuracy result. Moreover, the speedup
value is considerably increased when using single precision while the accuracy is maintained.

\item {\bf Robustness:} Our method achieves an accuracy similar to the classical Fast Marching
method. Additionally, it performs better when applied to irregular meshes and large meshes. 
\end{itemize}

From the experiments, we can conclude that our method, \appname, achieves competitive speedup
values without any preprocessing time. For problems where multiple sources are required and when
intensive distance queries will not be performed subsequently, such as the FPS algorithm, the
speedup increases, as the number of sources increases.

A limitation of our method is that it requires triangular meshes. Also,
our method requires more iterations when dealing with irregular meshes.

As future work, we aim at finding a tighter upper bound for the algorithm asymptotic complexity
than the one presented in our complexity analysis. Also, we plan to give a formal analysis
of the relation between $\epsilon$ and the obtained distance error.
 
Finally, we also want to investigate the use of our method as part of a method to compute
Centroidal Voronoi tessellations.

\section*{Acknowledgments}
The authors would like to express their gratitude to Cristian Lopez Del Alamo, Hueverton Souza, and Marcos Lage for their help and their valuable inputs.
 We also thank CAPES for funding this work. Models are courtesy of Stanford University (\textit{armadillo, bunny, asian dragon}) and AIM@SHAPE shape repository.


\bibliographystyle{cag-num-names}
\bibliography{paper}


\appendix

\section{Complexity: Farthest Point Sampling}\label{app:fps}
The FM algorithm has a complexity of $O(n \log n)$
similar to the Dijkstra algorithm. The complexity of the FPS algorithm without taking into
consideration the cost of calculating distances is $O(mn)$, where $m$ is the number of samples in
$S$. However, to compute a sub-sampling in a triangular mesh, we must compute the distance map with
the FM algorithm in each iteration. Hence the FPS algorithm complexity is $O(mn\log n)$. 

The number of operations $f(n)$ of the FPS algorithm using the \appabrv algorithm is

\begin{equation}\label{eq:fps_operations}
f(n) = \displaystyle\sum_{i = 1}^m \left(c_1\frac{n\sqrt{n}}{\sqrt{i}} + c_2n\right) 
\end{equation}
where $c_1$ and $c_2$ are constants. The terms inside the sum in Equation \ref{eq:fps_operations} 
represent the operations in each iteration: the \appabrv algorithm
(Algorithm \ref{algorithms/ptp}) which computes the geodesic distance map
(first term) and the selection of the vertex with the maximum distance from the $i$ sources
(second term). Equation \ref{eq:fps_p} reduces Equation \ref{eq:fps_operations}:

\begin{equation}
\displaystyle\sum_{i = 1}^m \left(c_1\frac{n\sqrt{n}}{\sqrt{i}} + c_2n\right) \,\leq\, 
\displaystyle\sum_{i = 1}^m (c_1 + c_2)\left(\frac{n\sqrt{n}}{\sqrt{i}}\right) \,=\, 
\displaystyle c \sum_{i = 1}^m \frac{n\sqrt{n}}{\sqrt{i}}
\end{equation}
where $c = c_1 + c_2$. We can conclude that:

\begin{equation}\label{eq:fps_p}
\displaystyle\sum_{i = 1}^m \left(c_1\frac{n\sqrt{n}}{\sqrt{i}} + c_2n\right) \,\leq\, 
\displaystyle c \sum_{i = 1}^m \frac{n\sqrt{n}}{\sqrt{i}} \,=\,
\displaystyle c n\sqrt{n}\sum_{i = 1}^m \frac{1}{\sqrt{i}}
\end{equation}.

It is possible to prove that:
\begin{equation}\label{eq:fps_proof}
\displaystyle \sum_{i = 1}^m \frac{1}{\sqrt{i}} = O\left(\sqrt{m}\right)
\end{equation}

\begin{proof}
We can prove that:
\begin{equation}\label{eq:fps_proof}
\displaystyle \sum_{i = 1}^m \frac{1}{\sqrt{i}} = O\left(\sqrt{m}\right)
\end{equation}
as a consequence of the following facts:

\begin{equation}\label{eq:fps_proof}
\displaystyle \sum_{i = 1}^m \frac{1}{\sqrt{i}} \leq \int_1^m \frac{1}{\sqrt{x}}dx = 2\sqrt{m} - 2
\end{equation}
hence, we can claim that:
\begin{equation}
\displaystyle \sum_{i = 1}^m \frac{1}{\sqrt{i}} = O\left(\sqrt{m}\right).
\end{equation}

\end{proof}

Thus, combining Equation \ref{eq:fps_p} with Equation \ref{eq:fps_proof} we obtain

\begin{equation}
\displaystyle cn\sqrt{n}\sum_{i = 1}^m \frac{1}{\sqrt{i}} = O\left(n\sqrt{n}\sqrt{m}\right) = O\left(\sqrt{m}n^{3/2}\right).
\end{equation}




\section{Sketch Proofs}\label{app:proofs}

In this section we analyse the behavior of the number of \toplesets and the number of required
iterations for certain classes of meshes. We show that the number of \toplesets in a monotonically
increasing mesh is $\rho \leq \sqrt{n}$ and is $\rho \leq n$ for a stationary mesh in the worst case,
where $n$ is the number of vertices; also we have established a relation between $\rho$ and the
maximum number of iterations $K$ to compute the distances map for special classes of meshes. The
definitions of monotonically increasing, decreasing and stationary meshes are established in the
Section \ref{sec:analysis}.

\mybimage{figures/toplesets_sorted_meshes}{Sorted \toplesets meshes distribution.}

\subsection{Theorem \ref{theorem:max_n_toplesets}}\label{proof:max_n_toplesets}
\theoremmaxntoplesets*

\begin{proof}
Let a triangular mesh $T = (V, E, F)$ and the \toplesets $V_r$, $r \in [1:\rho]$.
We can stablish that:
\begin{equation}\label{eq:proof:max_n_toplesets:diff}
|V_r| - |V_{r-1}| \geq c
\end{equation}
where $c \geq 1$ and $V_0 = \{s\}$ is the set containing the source vertex $s$. 
Without loss of generality we can choose a constant $c = 1$. Then we have
\begin{equation}
|V'_r| = |V'_{r-1}| + 1
\end{equation}
where $V'_r$ is a new increasing distribution of \toplesets
with the minimum difference such that $|V'_{r - 1}| \leq |V_r|$, $r \in [1:\rho'-1]$. Observe that
$\rho \leq \rho'$; then, we can solve the recurrence:
$|V'_r| = r$ for all $r \in [1:\rho' - 1]$. Now, by $\displaystyle\sum_{r = 0}^{\rho'-1} |V'_r| = n$
we have:
$$ 1 + \sum_{r = 1}^{\rho'-1} r = n $$
$$ 1 + \frac{\rho'(\rho' - 1)}{2} = n $$
$$ 2 + \rho'^2 - \rho' = 2n $$
$$ \rho' = O\left(\sqrt{n}\right) $$
because the fact that $\rho \leq \rho'$, we can claim that
$$ \rho = O\left(\sqrt{n}\right) $$.
\end{proof}

Figure \ref{figures/toplesets_sorted_meshes} shows the distribution of \toplesets sorted by the
number of vertices, and the functions $y = x$ (green line) and $y = 2x$ (blue line). We can observe
that the curve increases over the minimum increase constant $c = 1$, which is considerate in the
proofs. This plots confirm that the number of \toplesets $\rho \leq \sqrt{n}.$


\subsection{Theorem \ref{theorem:max_n_iter}}\label{proof:max_n_iter}

To prove Theorem \ref{theorem:max_n_iter}, we need the next lemmas:

\begin{lemma}\label{lemma:min_n_toplesets}
Let $T = (V, E, F)$ be a triangular mesh, and $s \in V$ a source vertex; the number of \toplesets
$\rho$ is $\Omega(\log n)$.
\end{lemma}

\begin{proof}

\image{figures/vtoplesets}{0.4}{Counting vertices at \topleset $V_r$.}

To count the number of vertices $|V_r|$ at \topleset $r \in [1:\rho]$, we first consider the number
of vertices in \topleset $V_r$ that must be connected to \topleset $V_{r-1}$. We can claim that it is
at least equal to degree $\deg(v)-3$, for each $v \in V_{r-1}$. The number $3$ in $\deg(v)-3$
accounts for the two mandatory vertices connecting neighbors in the same \topleset $V_{r-1}$ together
with the neighbor vertex at \topleset $V_{r-2}$. This is the maximum number of vertices that $V_r$
must have satisfying the constraints given by the degrees of the vertices $v \in V_{r-1}$
(see Figure \ref{figures/vtoplesets}). This is represented by Equation \ref{eq:count_v_toplesets}:
\begin{equation}\label{eq:count_v_toplesets}
|V_r| \leq \displaystyle\sum_{v \in V_{r-1}} (\deg(v) - 3) - |V_{r-1}|, \hspace{1cm} |V_0| = 1
\end{equation}
We use the maximum degree $\Delta_V = \max_{v \in V} \deg(v)$, to limit the number of vertices
at \topleset $r$:
$$|V_r| \leq \displaystyle\sum_{v \in V_{r-1}} \deg(v) - 3|V_{r-1}| - |V_{r-1}| \leq \displaystyle\sum_{v \in V_{r-1}} \Delta_V - 4|V_{r-1}|$$
$$|V_r| \leq \displaystyle\sum_{v \in V_{r-1}} \Delta_V - 4|V_{r-1}|$$
$$|V_r| \leq \Delta_V|V_{r-1}| - 4|V_{r-1}|$$
\begin{equation}\label{eq:recurr_v_toplesets}
|V_r| \leq (\Delta_V - 4)|V_{r-1}|.
\end{equation}
Let $b = \Delta_V - 4$. Solving recurrence \eqref{eq:recurr_v_toplesets} we obtain:
$$|V_r| \leq b^r, \hspace{1cm} |V_0| = 1$$
Now, knowing that $\displaystyle\sum_{r = 0}^{\rho-1} |V_r| = n$, where $n = |V|$ we have
$$\sum_{r = 0}^{\rho-1} |V_r| \leq \sum_{r = 0}^{\rho-1} b^r $$
$$n \leq \displaystyle\frac{b^{\rho} - 1}{b - 1} \leq b^{\rho} $$
$$\log_b n \leq \rho $$
therefore, as long as $b$ is limited, we can conclude that the number of \toplesets
$\rho$ is $\Omega(\log n)$.
\end{proof}

\begin{lemma}\label{lemma:max_n_iter}
$K_r \leq \left\lceil \frac{\Delta_v - 3}{2} \right\rceil (r - 1) + 1$
iterations compute the final distances of all vertices $v \in V_r$.
\end{lemma}

\def\remain{|\overline{V_{r}}|}

\begin{proof}
Let $K_r$ be the total number of iterations necessary to fix the final distance for all vertices
$v \in V_r$. Similarly, let $K_{r-1}$ be the total number of iterations necessary to fix the final
distances of the vertices $v' \in V_{r-1}$ in the previous \topleset.

We claim that $V_r$ requires at most the number of iterations of $V_{r-1}$ plus $i$ iterations,
that is:
\begin{equation}\label{eq:recurr_iter_kr}
\displaystyle K_r \leq K_{r-1} + i
\end{equation}
solving this recurrence, we obtain:
\begin{equation}
\displaystyle K_r \leq i (r-1) +1.
\end{equation}

In order to compute a superior bound to $i$, we analyze the behavior (see Table
\ref{table:iter_behavior}) of the number of remaining vertices $\remain$ in $V_r$ to have its
distances fixed at each iteration $i'$ and the total number of fixed distances in
$V_{r-1} \cup V_r$ (first column of Table \ref{table:iter_behavior}).
\begin{table}[!htb]
\begin{center}
\caption{Updated vertices in $V_r \cup V_{r-1}$ at beginning of iteration $i'$ and remaining
vertices $\remain$ at end of iteration $i'$.}\label{table:iter_behavior}
\small
\begin{tabular}{|r|l|l|}\hline
 $i'$ 	& $V_r \cup V_{r-1}$	& $\remain$													\\\hline
 $0$	& $|V_{r-1}|$			& $|V_r|$													\\
 $1$	& $2|V_{r-1}|$			& $|V_r| - |V_{r-1}|$										\\
 $2$	& $4|V_{r-1}|$			& $|V_r| - |V_{r-1}| - 2|V_{r-1}|$							\\
 $3$	& $6|V_{r-1}|$			& $|V_r| - |V_{r-1}| - 2|V_{r-1}| - 2|V_{r-1}|$				\\
 $4$	& $8|V_{r-1}|$			& $|V_r| - |V_{r-1}| - 2|V_{r-1}| - 2|V_{r-1}| - 2|V_{r-1}|$\\
 $\vdots$ & $\vdots$ & $\vdots$ \\
 $i$	& $2i|V_{r-1}|$			& $|V_r| - |V_{r-1}| - 2(i - 1)|V_{r - 1}|$		\\\hline
\end{tabular}
\end{center}
\end{table}

In Figure \ref{figures/vtoplesetsv2}, the circular icon corresponds to vertices in $V_{r-1}$ with
distances already fixed. The triangle icons indicate vertices with distances fixed at iteration
$i' = 1$. At $i^{'} = 1$ each edge in $V_{r-1}$ together with a vertex in $v \in V_r$ (triangle icon)
defines an updating triangle that is used to compute the final distance of $v$. Hence, there are
$|V_{r-1}|$ new fixed distances totalizing $2|V_{r-1}|$. 
For $i' > 1$, we have the double of edges that can be used to define updating triangles. Hence,
$2|V_{r - 1}|$ vertices have their distances fixed except for $i' = i$
which may fix less than $|V_{r - 1}|$ vertices as $\remain \geq 0$ (the number of remaining
distances to be fixed cannot be negative).

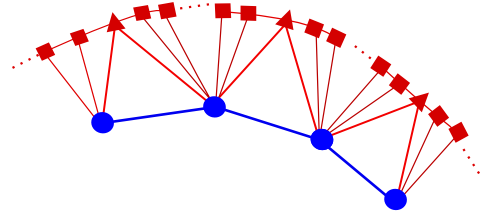
\image{figures/vtoplesetsv2}{0.7}{Counting iterations to updated the \topleset $V_r$.}

When all vertices $v \in V_r$ have their distances fixed, $\remain = 0$. Hence, we state that:
\begin{equation}
|V_r| - |V_{r - 1}| - 2(i - 1)|V_{r - 1}| = 0
\end{equation}
\begin{equation}
i = \left\lceil \frac{|V_r| + |V_{r - 1}|}{2|V_{r - 1}|} \right\rceil
\end{equation}

Consequently, we also claim that:
\begin{equation}
K_r \leq \left\lceil \frac{|V_r| + |V_{r - 1}|}{2|V_{r - 1}|} \right\rceil (r - 1) + 1
\end{equation}

Besides, according to Lemma \ref{lemma:min_n_toplesets}, equation \eqref{eq:recurr_v_toplesets}
(lower bound to the number of \toplesets of a mesh), we know that:
$$|V_r| \leq (\Delta_v - 4) |V_{r - 1}|$$
where $\Delta_v$ is the maximum degree of $v \in V_r$.

Thus,
$$|V_r| + |V_{r - 1}| \leq (\Delta_v - 4) |V_{r - 1}| + |V_{r - 1}|$$
$$\frac{|V_r| + |V_{r - 1}|}{2|V_{r - 1}|} \leq \frac{(\Delta_v - 4) |V_{r - 1}| + |V_{r - 1}|}{2|V_{r - 1}|}$$
$$\frac{|V_r| + |V_{r - 1}|}{2|V_{r - 1}|} \leq \frac{\Delta_v - 3 }{2}$$
finally:
$$K_r \leq \left\lceil \frac{|V_r| + |V_{r - 1}|}{2|V_{r - 1}|} \right\rceil (r - 1) + 1$$
\begin{equation}
K_r \leq \left\lceil \frac{\Delta_v - 3}{2} \right\rceil (r - 1) + 1.
\end{equation}

When the mesh is regular $\Delta_v = 6$ and we have:
$$K_r \leq 2(r - 1) + 1$$
$$K_r \leq 2r - 1.$$

When the mesh has \toplesets whose cardinality is stationary, that is $|V_r| = |V_{r - 1}|$ for all
$r$ we have:
$$K_r \leq \left\lceil \frac{|V_r| + |V_{r - 1}|}{2|V_{r - 1}|} \right\rceil (r - 1) + 1$$
$$K_r \leq \left\lceil \frac{2|V_{r - 1}|}{2|V_{r - 1}|} \right\rceil (r - 1) + 1$$
then $K_r \leq r$, and one iteration is sufficient to fix all distances in $V_r$ once $V_{r-1}$
has its distances fixed.

\end{proof}

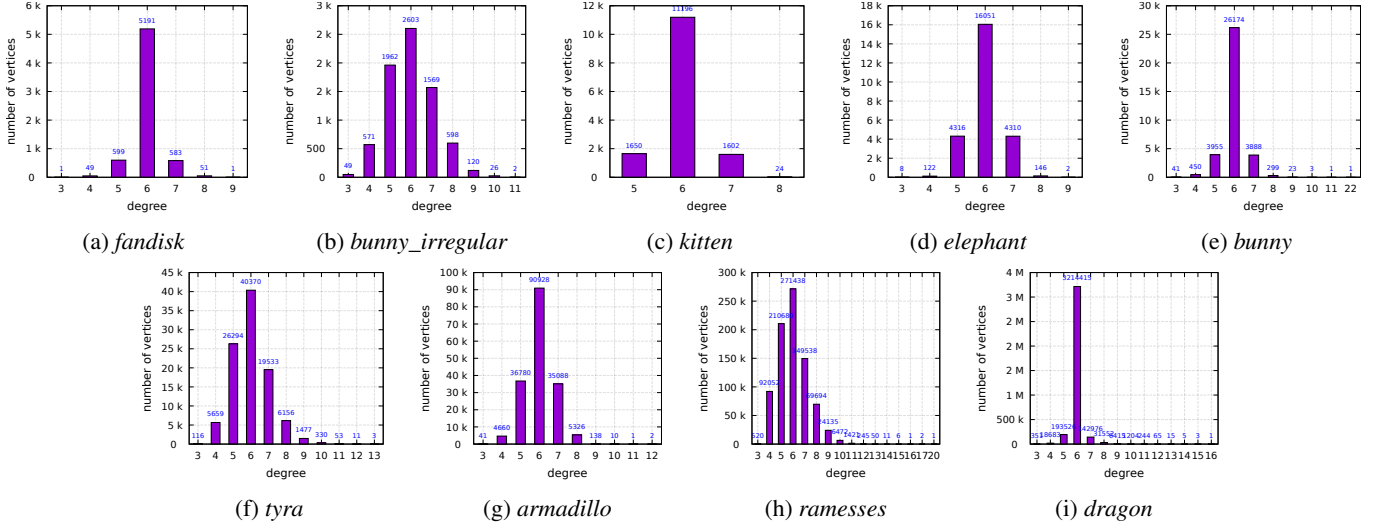
\mybimage{figures/deg_meshes}{Degree histogram for the meshes in the experiments.}

Figure \ref{figures/deg_meshes} shows the degree histogram considering all the meshes
used in the experiments. We can observe that the predominant degrees are 6, 4 and 5.


\theoremmaxniter*

\begin{proof}
To proof Theorem \ref{theorem:max_n_iter} we use the Lemma \ref{lemma:max_n_iter}. According
to it, the number of iterations $K$ to compute the final distance to the last \topleset
$V_{\rho}$ is $K = c(\rho - 1) + 1$,  $\rho \leq \sqrt{n}$, and
$c = \left\lceil \frac{\Delta_v - 3}{2} \right\rceil $. Thus, we can claim that $K \leq c\sqrt{n}$
and $K = O\left(\sqrt{n}\right)$. Observe that for regular meshes $c = 2$.
\end{proof}

\end{document}